\documentclass[aps,prx,floatfix,twocolumn,superscriptaddress,10pt]{revtex4-2}

\usepackage[utf8]{inputenc}
\usepackage[T1]{fontenc}
\usepackage{lmodern}
\usepackage{orcidlink}
\usepackage{soul}
\usepackage{color}
\usepackage[section]{placeins}
\usepackage{latexsym}
\usepackage{graphicx}
\usepackage{rotating}
\usepackage[normalem]{ulem}
\usepackage{hyperref}
\usepackage{amsmath,amsfonts}
\usepackage{bm}
\usepackage{color}
\usepackage{float}
\usepackage{textgreek}
\usepackage{qcircuit}
\usepackage{listings}
\usepackage[greek,english]{babel}
\usepackage{mathrsfs}
\usepackage{physics}
\usepackage{braket}
\usepackage{algorithm}
\usepackage{algpseudocode}
\usepackage{multirow}

\hypersetup{colorlinks=true,urlcolor=blue,linkcolor=blue}

\newcommand{\EE}{\mathbb{E}}
\newcommand{\cN}{\mathcal{N}}

\renewcommand{\d}{\mathrm{d}}

\newcommand{\mt}{G}
\newcommand{\force}{{\bm b}}
\newcommand{\setH}{S_H}
\newcommand{\setNN}{S_{\mathrm{NN}}}
\newcommand{\setTFIM}{S_{\mathrm{IM}}}

\DeclareMathOperator{\Id}{Id}
\DeclareMathOperator{\Var}{Var}
\DeclareMathOperator{\Covar}{Covar}

\let\vec\bm

\def\d{\delta}

\let\originalpolishl\l
\def\l{\lambda}
\let\redefinedl\l

\begin{document}

\title{Operator-Projected Variational Quantum Imaginary Time Evolution}
\author{Aeishah Ameera Anuar\orcidlink{0009-0006-3228-3086}}
\email{aeishah.ameera@meetiqm.com}
\affiliation{IQM Quantum Computers France SaS, 40 Rue du Louvre, 75001 Paris, France}
\affiliation{Sorbonne Universit\'e, CNRS, Laboratoire de Physique Th\'eorique de la Mati\`ere Condens\'ee, LPTMC, F-75005 Paris, France}
\author{François Jamet\orcidlink{0000-0002-5679-209X}}
\affiliation{IQM Quantum Computers France SaS, 40 Rue du Louvre, 75001 Paris, France}
\author{Fabio Gironella\orcidlink{0000-0001-6922-7340}}
\affiliation{CNRS - Laboratoire de Math\'ematiques Jean Leray, Universit\'e de Nantes, France}
\author{Fedor Šimkovic IV\orcidlink{0000-0003-0637-5244}}
\email{fedor.simkovic@meetiqm.com}
\affiliation{IQM Quantum Computers, Georg-Brauchle-Ring 23-25, 80992, Munich, Germany}
\author{Riccardo Rossi\orcidlink{0000-0003-1384-6235}}
\affiliation{Sorbonne Universit\'e, CNRS, Laboratoire de Physique Th\'eorique de la Mati\`ere Condens\'ee, LPTMC, F-75005 Paris, France}
\affiliation{Institute of Physics, \'Ecole Polytechnique F\'ed\'erale de Lausanne (EPFL), CH-1015 Lausanne, Switzerland}

\begin{abstract}
Variational Quantum Imaginary Time Evolution (VQITE) is a leading technique for ground state preparation on quantum computers.  A significant computational challenge of VQITE is the determination of the quantum geometric tensor. We show that requiring the imaginary-time evolution to be correct only when projected onto a chosen set of operators allows to achieve a twofold reduction in circuit depth by bypassing fidelity estimations, and reduces measurement complexity from quadratic to linear in the number of parameters. We demonstrate by a simulation of the transverse-field Ising model that our algorithm achieves a several orders of magnitude improvement in the number of measurements required for the same accuracy.
\end{abstract}

\maketitle

\section{INTRODUCTION}\label{sec:intro}

Investigating ground states and non-equilibrium dynamics of strongly correlated systems is key to predicting the quantum properties of real materials with immediate industrial relevance for various applications such as battery design, solar cells or nitrogen fixation~\cite{Kim_2022}. For decades, many numerical tools~\cite{qmc, mps, dmrg} have been developed for this purpose, yet their predictive power is fundamentally limited by the exponential growth of the Hilbert space with system size and the fermionic nature of the problems of interest, resulting in the infamous fermionic sign problem~\cite{sign}. Many physically interesting systems thus fall outside the reach of current numerical methods, exposing the need for the development of alternate approaches. 

This has been a prominent motivational factor behind the development of  quantum computers, following the ideas that quantum systems can be efficiently represented by another quantum system and that real time evolution of the Schrödinger equation can be realized via unitary quantum circuits~\cite{feynman}. Yet, the enthusiasm over a theoretical quantum superiority~\cite{bqp} has been offset by the extreme fragility of quantum states in current noisy intermediate-scale quantum devices (NISQ)~\cite{Preskill}, thus limiting quantum algorithms to short circuits and low operation counts. 

Variational quantum algorithms were found to work well within these hardware constraints and are leading contenders in the race to practical quantum advantage \cite{hybrid1, qaoa, hybrid3, hybrid4,McClean_2016}. One well-studied example is the Variational Quantum Eigensolver (VQE) which leverages the quantum computer to prepare a parameterised trial state and to measure the corresponding energy, which is then minimized by a classical optimizer in an external loop \cite{vqe,vqe1, vqe-chem, vqe-chem1, vqe-chem2, vqe-ss, vqe-ss1}. VQE is modular, flexible, and straightforwardly implemented, but this comes at the price of a lack of performance guarantees. 
Indeed, recent works have identified practical as well as fundamental challenges for VQE related to measurement, overparametrization, and \emph{barren plateau} optimisation landscapes~\cite{meas, bp1, bp2, bp3}. 

As an alternative, several quantum algorithms based on imaginary time evolution (ITE) \cite{ite} have been proposed to prepare ground states of many-body systems by systematically suppressing excited states. The earliest, Quantum Imaginary Time Evolution (QITE) approach approximates the Trotterized imaginary time evolution using unitary transformations which are executable on a quantum computer \cite{Motta_2019}. While effective, QITE can be computationally expensive due to the requirement for precise gate operations over many Trotter steps, even if multiple variants aimed at improving these bottlenecks have been introduced \cite{mqite,pqite,v-tay-qite,adia-qite,Mcardle}.

Variational Quantum Imaginary Time Evolution, (VQITE) based on McLachlan's variational principle is a more NISQ-friendly approach to QITE. This method deterministically evolves parameters in order to minimize the McLachlan distance, a measure of how well the variational state approximates the exact propagation of a quantum state along the imaginary time axis \cite{Mcardle, mclahans, Gacon_2024, qng, adapt-qite}. The method can be generalized to real-time evolution, where the Dirac-Frenkel variational principle is used instead~\cite{mclahans}. Both of these algorithms necessitate the calculation of the Quantum Fisher Information matrix (QFI), a process that results in a prohibitively large number of  circuit evaluations for near-term quantum devices \cite{qfi, qng, grad1, yamamoto2019naturalgradientvariationalquantum}.

To address this issue, Ref.~\cite {Gacon_2023} introduced a stochastic evaluation of the QFI,  allowing for a trade-off between resource requirements and accuracy. The practical near-term suitability of this algorithm has been showcased by a 27 qubit simulation of the tranverse field Ising model. The DualQITE algorithm is another proposed alternative, where the calculation of the QFI is replaced by a dual problem requiring fidelity-based optimization \cite{Gacon_2024}, which improves the measurement complexity from quadratic to linear in the number of parameters.

In this paper, we introduce the Operator-Projected Variational Quantum Imaginary Time Evolution algorithm (OVQITE), an approach that aims at variationally reproducing the ITE on a set of chosen observables. We show that OVQITE halves the quantum circuit depth of VQITE by avoiding fidelity calculations, and allows for a linear measurement complexity scaling as a function of the number of parameters, thus paving the way for more practical implementations of quantum simulations on NISQ devices.
By benchmarking on the transverse-field Ising model, we show that OVQITE allows for a substantial reduction in the number of measurements to reach a given accuracy when compared to VQITE.

The structure of the paper goes as follows: In Section~\ref{sec:method}, we provide the background on imaginary-time evolution and introduce our new approach, OVQITE. In Section~\ref{sec:scaling}, we discuss the quantum resources required for both the VQITE and OVQITE algorithms. We numerically investigate the TFIM and analyze the convergence properties of both approaches in Section~\ref{sec:results} before presenting concluding remarks in Section~\ref{sec:conclusions}.

\section{Theory}\label{sec:method}
In this section, we present the details of the OVQITE technique introduced in this work in a general setting. We start with a summary of the basic notions of imaginary-time evolution (ITE) in Sec~\ref{subsec:ite}. Then, we introduce the operator-projected ITE (OITE) in Sec.~\ref{subsec:oite} and discuss a quantum variational implementation of the OITE equations in Sec.~\ref{subsec:ovqite}.

\subsection{Imaginary Time Evolution}\label{subsec:ite}
Let $\hat \rho$ be the density matrix of a quantum system of $n$ qubits. From $\hat \rho$, the expectation value of an operator $\hat{O}$ can be computed as:
\begin{equation}\label{eq-expansion-operators-dm}
    \braket{\hat{O}}_{\hat{\rho}} \equiv \mathrm{Tr}\left[\hat{\rho}\,\hat{O}\right],
\end{equation}
where we assume that $\mathrm{Tr}\,\hat{\rho} = 1$. Let $\mathcal{B}=\{\hat O_1,\hat O_2,\dots\}$ be a complete basis of operators acting on the $n$-qubit Hilbert space. By definition, the density matrix $\hat \rho$ can be expanded in the $\mathcal{B}$ basis:
\begin{equation}\label{eq-expansion-operators-dm2}
    \hat \rho = \sum_{j} \rho^{(j)}\,\hat O_j,
\end{equation}
where $\rho^{(j)}\in\mathbb{C}$. Let us denote the Hilbert-Schmidt inner product of operators by $\braket{\hat A,\hat B}_{\text{HS}}\equiv\text{Tr}\,\hat A^\dagger \hat B$. If the basis $\mathcal{B}$ is orthogonal, i.e. $\braket{\hat O_j,\hat O_k}_{\text{HS}} = \delta_{j,k}\braket{\hat O_j,\hat O_j}_{\text{HS}}$, one can write $\rho^{(j)}= \frac{\braket{\hat O_j}_{\hat\rho} ^*}{\braket{\hat O_j,\hat O_j}_{\text{HS}}}$.

For a time-independent Hamiltonian $\hat H$, the imaginary time evolution (ITE) is characterized by a non-unitary operator $e^{-\tau \hat H}$, where $\tau \ge 0$. The imaginary-time evolution of a density matrix of an initial state $\hat\rho_0$ up to imaginary time $\tau$ is given by:
    \begin{equation}
    \label{eqn:ite}
        \hat \rho_\tau = \frac{e^{-\tau\hat H}\,\hat\rho_0\, e^{-\tau\hat H}}{\text{Tr}\big[ e^{-2 \tau \hat H }\hat\rho_0]}.
    \end{equation}
Imaginary time evolution can be used to prepare thermal states and, in particular, the ground state of a quantum system:
    \begin{equation}
    \braket{\hat{H}}_{\hat\rho_{\tau}}\overset{\tau\to\infty}{=} E_0  
    \end{equation}
where $E_0$ is the ground state energy and it is assumed that $\braket{\psi|\hat{\rho}_0|\psi}\neq 0$ for at least one eigenstate $\ket{\psi}$ of $\hat{H}$ with eigenvalue $E_0$. In this work, we focus exclusively on ground state preparation.
    
 We can rewrite Eq.~\eqref{eqn:ite} in a differential form: 
    \begin{equation}\label{eq-ite}
        {\partial_\tau \hat\rho_\tau} = -\{\hat H,\hat\rho_\tau\}+2\braket{ \hat H}_{\hat\rho_\tau}\, \hat\rho_\tau,
    \end{equation}
with the anticommutator $\{\hat A,\hat B\} \equiv \hat A\hat B+\hat B\hat A$ and using the simplified notation of $\partial_\tau \equiv \partial / \partial\tau$. 
Let us consider an expansion of $\hat \rho_\tau$ in a $\mathcal{B}$ operator basis as defined by Eq.~\eqref{eq-expansion-operators-dm2}, with $\tau$-dependent coefficients $\rho_\tau^{(j)}$. This allows for Eq.~\eqref{eq-ite} to be reformulated as:
    \begin{equation}\label{eq-ite-operators}
    \begin{split}
       \partial_\tau \rho_\tau^{(j)}&=-\sum_{k,l}\tilde{G}^{-1}_{j,k}\braket{\{\hat{O}_k,\hat{O}_l^\dagger\}, \hat H }_{\text{HS}}\,\rho_\tau^{(l)}+\\
       &+2\braket{\hat H}_{\hat \rho_\tau}\rho_\tau^{(j)},
       \end{split}
    \end{equation}
where we have supposed that the matrix $\tilde{G}_{j,k}\equiv\braket{\hat O_j,\hat O_k}_{\text{HS}}$ is invertible.

\subsection{Operator-projected ITE (OITE)}\label{subsec:oite}
The ITE formulation of Eq.~\eqref{eq-ite-operators} involves an expansion over a basis consisting of an exponential number of operators acting on an $n$-qubit system, which therefore prevents a direct practical implementation. In this work, we focus on using ITE to prepare the ground state of 2-local Hamiltonians, which is fully characterized by the expectation value of 2-local operators~\cite{coleman1963structure}. This observation is at the basis of the reduced density matrix (RDM) numerical method~\cite{cioslowski2012many,coleman2007reduced}, where one minimizes the energy of an RDM with constraints coming from the N-representability, which makes the problem QMA-complete~\cite{liu2007quantum}. 
The correspondence between expectation values of 2-local operators and ground state density matrices suggests that, if $\hat{\rho}_\tau$ is sufficiently close to a ground state of a 2-local Hamiltonian, the set of equations in Eq.~\eqref{eq-ite-operators} is effectively redundant as expectation values of 2-local operators are sufficient to characterize the quantum state~\footnote{We note that our method can in principle be also applied to arbitrary $k$-local Hamiltonians.}.

Another motivation for the effective restriction of the operator set which we use in Eq.~\eqref{eq-ite-operators} comes from a property of NISQ quantum hardware: typically, the expectation value of $k$-local operators will decay exponentially fast with $k$ in the presence of quantum noise. This means that resolving the expectation value of $k$-local operators on real hardware with fixed relative precision is expected to be challenging for large $k$, while at the same time the magnitude of these expectation values is usually small. 

Motivated by these arguments, let us rewrite Eq.~\eqref{eq-ite-operators} for a set of operators $S_\tau$ whose cardinality increases at most polynomially with the number of qubits, $|S_\tau|=O(\mathrm{poly}(n))$, to obtain the Ehrenfest theorem formulated for operators from the set, $\hat O\in S_\tau$:
\begin{equation}\label{eq-ite-projected-operators}
    \begin{split}
       \partial_\tau \braket{\hat O}_{\hat\eta_\tau}       =-\braket{\{\hat{H},\hat{O}\}}_{\hat\eta_\tau}+2\braket{\hat H}_{\hat\eta_\tau} \braket{\hat O}_{\hat\eta_\tau},
       \end{split}
    \end{equation}
where $\hat \eta_\tau$ is a density matrix at imaginary time $\tau \ge 0$. We stress that Eq.~\eqref{eq-ite-projected-operators} describes the time evolution of a family of valid density matrices $\{\hat\eta_\tau\}$, meaning they are Hermitian and positive semi-definite operators. The elements of $\hat \eta_\tau$ coincide with $\hat \rho_\tau$ of Eq.~\eqref{eqn:ite} if $S_{\tau'}$  is an operator basis for all intermediate imaginary times $\tau'$ such that ${0\le \tau'\le \tau}$. Let us also remark that our method avoids the N-representability issue of the RDM numerical method as we explicitly derive our RDM from a quantum state. We also point out that while one of the motivations for OITE was choosing the RDM to coincide with the $S_\tau$ operator set, we are not limited to this choice, as we will show in the implementation part of this manuscript.

\subsection{Operator-projected Variational Quantum ITE (OVQITE)}\label{subsec:ovqite}
Let $\hat \rho_{\bm \theta}$ be the density matrix of a parameterized differentiable quantum circuit with parameters $\bm \theta\in\mathbb{R}^{N_\theta}$, where $N_\theta$ is the number of parameters in the circuit. Starting from some initial values for the parameters $\bm \theta_{\tau=0}$, we would like to build a sequence of parameter values $\bm \theta_{\tau = u \delta}$, with $u\in\{1,2,\dots\}$ and $\delta>0$, such that $\hat  \rho_{\bm\theta_{\tau=u\delta}}$ approximates the operator-projected ITE of Eq.~\eqref{eq-ite-projected-operators}. We introduce a loss function $L$ that tracks the variational error in Eq.~\eqref{eq-ite-projected-operators} as a function of the imaginary time derivative of the parameters $\dot{\bm \theta}\equiv\partial_\tau \bm \theta$:
\begin{equation}\label{eq-oite-loss}
\begin{split}
    &L(\dot{\bm \theta}|\bm \theta,S_\tau)\equiv\frac{1}{2}\sum_{\hat O\in S_\tau}\left|\dot {\bm \theta} \cdot \nabla_{\bm \theta} \braket{\hat O}_{\hat \rho_{\bm \theta}}-V(\hat O, \hat{ \rho}_{\bm \theta})\right|^2=\\
    &=\frac{1}{2}\dot {\bm \theta}\cdot \mt_{\bm \theta,S_\tau} \,\dot {\bm \theta} -\dot{\bm \theta}\cdot \force_{\bm \theta,S_\tau}+\frac{1}{2}\sum_{\hat O\in S_\tau}|V(\hat O,\hat \rho_{\bm \theta})|^2,
    \end{split}
\end{equation}
with 
\begin{equation}\label{eqn:V}
V(\hat O,\hat \rho_{\bm \theta})\equiv-\braket{\{\hat{H},\hat{O}\}}_{\hat \rho_{\bm \theta}}+2\braket{\hat H}_{\hat \rho_{\bm \theta}} \braket{\hat O}_{\hat \rho_{\bm \theta}},
\end{equation}
\begin{equation}\label{eqn:obs_matrix}
   [ \mt_{\bm \theta,S_\tau}]_{j,k} \equiv\mathrm{Re}\sum_{\hat O\in S_\tau}\left[\partial_{\theta_j}\braket{\hat O}_{\hat \rho_{\bm \theta}}\right]^*\partial_{\theta_k}\braket{\hat O}_{\hat \rho_{\bm \theta}},
\end{equation}
\begin{equation}\label{eqn:target_vec}
\force_{\bm \theta,S_\tau}\equiv\mathrm{Re}\sum_{\hat O\in S_\tau}\left[\nabla_{\bm \theta}\braket{\hat{O}}_{\hat \rho_{\bm \theta}}\right]^*\,V(\hat O,\hat \rho_{\bm \theta}).
\end{equation}
We remark that $\mt_{\bm \theta,S_\tau}$ is an $N_\theta\times N_\theta$ positive semi-definite real matrix of rank upper bounded by $|S_\tau|$.
Eq.~\eqref{eq-oite-loss} is formally identical to the VQITE loss function (see e.g. Ref.~\cite{mclahans}) with the substitution of $\mt_{\bm \theta,S_\tau}\to \mt_{\bm \theta}^{(\mathrm{VQITE})}$ and $\force_{\bm \theta,S_\tau}\to \force_{\bm \theta}^{(\mathrm{VQITE})}$, given by:
\begin{equation}
\begin{split}
    &[\mt_{\bm \theta}^{(\mathrm{VQITE})}]_{j,k}\equiv\\
&\quad\text{Re}\left[\braket{\partial_{\theta_j}\psi_{\bm \theta}|\partial_{\theta_k}\psi_{\bm \theta}}-\braket{\partial_{\theta_j}\psi_{\bm \theta}|\psi_{\bm \theta}}\braket{\psi_{\bm \theta}|\partial_{\theta_k}\psi_{\bm \theta}}\right],
    \end{split}
\end{equation}
\begin{equation}\label{eqn:f}
    \force_{\bm \theta}^{(\mathrm{VQITE})}\equiv-\nabla_{\bm \theta}\braket{\psi_{\bm \theta}|\hat H|\psi_{\bm \theta}},
\end{equation}
where $\ket{\psi_{\bm \theta}}$ is a parameterized quantum circuit ansatz for the wavefunction.
  
The values of $\dot{\bm \theta}$ that minimize the loss function of Eq.~\eqref{eq-oite-loss} satisfy the following equation:
\begin{equation}\label{eq-s-matrix-and-f}
    \mt_{\bm \theta,S_\tau}\dot{\bm \theta}=\force_{\bm \theta,S_\tau}.
\end{equation}
At each step of OVQITE, the variational parameters of the quantum circuit are updated as $\bm \theta_{\tau +\delta} =\bm \theta_\tau+\delta\,\bm \dot{\bm \theta}$, for a small $\delta>0$.

In the presence of shot noise due to the finite number of measurements available on quantum hardware, or when the rank of the $\mt_{\bm \theta, S_\tau}$ is not maximal, a regularization procedure is required in order to reliably evaluate Eq.~\eqref{eq-s-matrix-and-f}, see Appendix~\ref{sec:pinv}. An optimal algorithm to solve the noisy linear problem implied by Eq.~\eqref{eq-oite-loss} in the realistic case of shot noise induced bias in $\mt_{\bm \theta,S_\tau}$ and $\force_{\bm \theta,S_\tau}$ is detailed in Appendix~\ref{sec:error-in-variables}, although its numerical implementation is left to future work. When $S_\tau$ is chosen to be the set of all adjacent Pauli strings on $n$ qubits, the algorithm shares some similarities with the QITE approach of Ref.~\cite{Motta_2019}, with the important difference that we optimize a fixed quantum circuit instead of adding a new layer at each imaginary-time step.

\section{Quantum Resource Estimate}\label{sec:scaling}
In this section, we show how to implement the OVQITE algorithm on a quantum device, and investigate the scaling with respect to the number of terms in the Hamiltonian of interest, $N_{H}$, as well as the number of parameters in the variational quantum circuit used for state preparation, $N_{\theta}$. Additionally, we study analytically and numerically the influence of shot noise on the quality of results, i.e. the sampling errors from from a finite number of measurements, $N_{\text{sh}}$. Due to the prohibitive classical computational cost of simulating noisy quantum circuits of relevant size, quantum hardware noise is not considered here and instead left to future realizations on real quantum processors. We first investigate the circuit depth after state preparation and then discuss the total number of calls required to implement OVQITE and compare against VQITE. The summary of our findings can be found in Table~\ref{table}.



\subsection{Quantum implementation of OVQITE}
We specialize the discussion to a noiseless parameterized quantum state of the form $\hat{\rho}_{\bm \theta}:=\ket{\psi_{\bm \theta}}\bra{\psi_{\bm \theta}}=U(\bm \theta)\ket{\bm 0}\bra{\bm 0}U^\dagger(\bm \theta)$, where $\bm \theta\in\mathbb{R}^{N_\theta}$.
For OVQITE, the observation matrix of Eq.~\eqref{eq-s-matrix-and-f} can be written as the product of two rectangular matrices
\begin{equation}\label{eq-G-from-M}
    \mt_{\bm \theta,S_\tau}=\mathrm{Re}\;M^\dagger M,
\end{equation}
where $M\in \mathbb{R}^{|S_\tau|\times N_\theta}$ with matrix elements given by the derivatives of expectation values of operators from the set  $S_\tau=\{\hat O_i\}_{i=1}^{|S_\tau|}$, such that:
	\begin{equation}
	\label{eqn:psr0}
	\begin{aligned}
M_{i,j}=\partial_{\theta_j}\braket{ \hat O_i}_{\bm\theta},
		\end{aligned}
	\end{equation}
 where we have used the shorthand notation $\braket{\hat O}_{\bm \theta}\equiv \braket{\hat O}_{\hat{\rho}_{\bm \theta}}$.  We further introduce 
 \begin{equation}\label{eqn:v-the-vector}
     v_j \equiv V(\hat{O}_j, S_\tau),
 \end{equation}
 for ${j\in\{1,\dots,|S_\tau|\}}$, and we have
 \begin{equation}\label{eq-b-from-M-v}
     \bm b_{\bm \theta,S_\tau} = \mathrm{Re}\;M^\dagger \bm v.
 \end{equation}
 The matrix $M$ of Eq.~\eqref{eqn:psr0} can be evaluated using the parameter-shift rule (PSR), which gives an analytical gradient formula for parameterized quantum circuits that consist of tensor products of Pauli rotation gates~\cite{grad1,grad2}. Thus, at each step, populating the matrix $\mt_{\bm \theta,S_\tau}$ amounts to measuring: 
\begin{equation}
\label{eqn:psr}
\begin{aligned}
M_{i,j}={1\over2 \sin s}&\big(\braket{ \hat O_i}_{\bm\theta+s\vec e_j}-\braket{ \hat O_i}_{\bm\theta-s\vec e_j}\big),
\end{aligned}
\end{equation}
with $s$ being a constant with $s\neq k\pi$ and $\vec e_j$ being the unit vector along the $\theta_j$ axis. This task requires only the evaluation of the expectation value for two-local Pauli strings at two shifted parameter values. An additional advantage of using the PSR is that has also been shown to possess a degree of inherent noise-resilience~\cite{meyer2021variational, piskor2022using}, making it thus suitable for NISQ applications.

  
We further note that $V(\hat O,\hat  \rho_\theta)$ of Eq.~\eqref{eqn:target_vec} is a function of the expectation values of a set of operators. For its evaluation, the algorithm requires the addition of a maximum of two layers of single qubit gates after the state preparation circuit in order to measure arbitrary operators in the computational basis.

\subsection{Scaling analysis}
\label{sec-scaling-analysis}
Let us now investigate the quantum cost of the OVQITE algorithm. We note that we will not discuss here the classical computational cost, which is polynomial in the number of circuit parameters. The total number of quantum measurements at each step, $\mathcal{M}$, is obtained as
 \begin{equation}\label{eqn-shots}
 \mathcal{M}  = N_{\mathrm{sh}}\,N_{\mathrm{circ}},
 \end{equation}
where $N_{\mathrm{sh}}$ is the number of shots, and   $N_{\mathrm{circ}}$ is the number of quantum circuits to be measured. In this section we focus on the estimation of $N_{\mathrm{circ}}$. We assume here that each operator $\hat{O}\in S_\tau$ consists of a single Pauli string in order to simplify the discussion. Let us first describe a naive measurement schedule, and then later discuss an optimized version of it. At each iteration of OVQITE, the total number of circuits $N_{\mathrm{circ}}$ required to populate the 
matrix $M$ of Eq.~\eqref{eqn:psr0} scales as $\mathcal{O} (N_\theta\, |S_\tau|  )$. The vector $\bm v\in \mathbb{R}^{|S_\tau|}$ of Eq.~\eqref{eqn:v-the-vector} requires a total of ${\mathcal{O}(\sum_j N_{\{\hat H, \hat O_j\}}+|S_\tau|+N_{\hat H})}$ circuits, where we have defined $N_{\hat{O}}$ as the number of Pauli strings in the decomposition of the operator $\hat{O}$. For general local Hamiltonians, the number of Pauli strings in the anticommutator of the Hamiltonian $\hat H$ with a Pauli string operator $\hat O$ is defined as
\begin{equation}
    N_{\{\hat H,\hat O\}} 
    = \mathcal{O}(N_{\hat H}|S_\tau|),
\end{equation} 
since for each operator one needs to compute the expectation value of the anticommutator involving all terms of the Hamiltonian. Under these assumptions, the leading scaling of the number of circuit is $\mathcal{O}(N_{\hat H} \,|S_\tau|)$. For systems with finite-ranged interactions, $N_{\hat H}$ scales linearly with the number of qubits $n$, which implies $N_{\mathrm{circ}}=\mathcal{O}(n\, |S_\tau|)$, while for more complex systems, such as those from quantum chemistry, the scaling of $N_{\hat H}$ with the number of qubits can be higher. \\

It is possible to mitigate the measurement overhead by adopting optimization strategies, such as classical shadows protocols and Pauli grouping schemes, which have been previously shown to be particularly effective for this task~\cite{meas,meas1,meas2,vqe1}. In particular, grouping strategies can reduce the scaling dependencies from $N_H$ and $|S_\tau|$ to $C_{N_H}$ and $C_{|S_\tau|}$ where where $C_X$ is the total number of counts of distinct groups contained in $X$ where all members can be measured simultaneously due to their commutation properties. Thus, we have that $C_X\leq |X|$. The scaling with $N_H$ maybe be further reduced through aforementioned grouping strategies, as summarized in Table~\ref{table} which compares a naive individual measurement strategy, $N^{(\mathrm{naive})}_{\mathrm{circ}}$, with an optimized, grouping based simultaneous measurement strategy, $N^{(\mathrm{opt})}_{\mathrm{circ}}$.

Let us now focus on the scaling comparison between OVQITE and VQITE. In contrast to OVQITE, the implementation of VQITE  requires the evaluation of a metric tensor which consists of overlaps of states with different parameter values.  The gradient of the energy in Eq.~\eqref{eqn:f} can be obtained by preparing  $\mathcal{O}(N_\theta \,N_{\hat H})$ circuits.
Specifically, the real part of the quantum geometric tensor, $\mt_{\bm \theta}^{(\mathrm{VQITE})}$, takes the form of a Hessian matrix, with its entries being the second order partial derivatives of an expectation value called the survival probability~\cite{qfi,grad1}: 
\begin{equation}
\label{qgt}
    [\mt_{\bm \theta}^{(\mathrm{VQITE})}]_{ij}=-{1 \over 2}{\partial^2 \over \partial\theta_i\partial \theta_j}|\braket{\psi_{\bm\theta}|\psi_{\bm\theta'}}|^2\bigg|_{\bm\theta'=\bm\theta}
\end{equation}

Each matrix entry can be evaluated using the parameter shift rule twice, as described by the following expression: 
\begin{equation}
\label{eqn:overlap}
\begin{aligned}
    [\mt_{\bm \theta}^{(\mathrm{VQITE})}]_{ij}=-{1\over 8} & \big(\;|\braket{\psi_{\bm \theta}|\psi_{\bm\theta+s(\vec e_i + \vec e_j)}}|^2\\
    &-|\braket{\psi_{\bm\theta}|\psi_{\bm\theta+s(\vec e_i - \vec e_j)}}|^2\\
    &-|\braket{\psi_{\bm\theta}|\psi_{\bm\theta-s(\vec e_i - \vec e_j)}}|^2\\
    &+ |\braket{\psi_{\bm\theta}|\psi_{\bm\theta-s(\vec e_i + \vec e_j)}}|^2\big)
    \end{aligned}
\end{equation}

The overlaps in the expression above can be computed using the SWAP-test, or by applying the conjugate of the state preparation unitary $U(\bm\theta)$ (defined by $\ket{\psi_{\bm \theta}} \equiv U(\bm\theta) \ket{\mathbf{0}}$) at shifted parameters \cite{grad1} $U^{\dagger}(\bm\theta+s(\vec e_i - \vec e_j))$ and measuring the all-zero bit-string in the computational basis state, i.e. the overlap $|\braket{\psi_{\bm\theta}|\psi_{\bm\theta+s(\vec e_i - \vec e_j)}}|^2$ is the probability of observing the outcome 0 for the circuit $U^{\dagger}(\theta+s(\vec e_i - \vec e_j)) U(\theta) \ket{\mathbf{0}}$. The resulting circuit for the evaluation of matrix elements of $\mt_{\bm \theta}^{(\mathrm{VQITE})}$ thus has  twice the circuit depth of the corresponding state preparation and hence roughly twice the depth of the circuits used in OVQITE. The key observation is that the circuit complexity of computing overlaps needed for VQITE is reduced to the circuit complexity of computing expectation values in the OVQITE algorithm. It can be shown that the number of circuits to be prepared, $N_{\mathrm{circ}}$, needed to estimate $\mt_{\bm \theta}^{(\mathrm{VQITE})}$ and $\force_{\bm \theta}^{(\mathrm{VQITE})}$ scales as $\mathcal{O}(N_\theta^2)$ and $\mathcal{O}(N_\theta N_H)$, respectively~\cite{grad1,Mitarai_2019}. 

From Table~\ref{table} we can see that the VQITE algorithm scales quadratically with the number of tunable parameters $N_\theta$, in the large $N_{\theta}$ limit.  This is in contrast with OVQITE, which scales linearly with $N_{\theta}$, with a multiplicative overhead of $C_{S_{\tau}}$.

\begin{table}[h!]
\begin{tabular}{ c c c c } 
\hline
Algorithm & Quantity & $N^{(\mathrm{naive})}_{\mathrm{circ}}$ & $N^{(\mathrm{opt})}_{\mathrm{circ}}$ \\
\hline
\multirow{3}{4em}{VQITE } &  &  \\
& $\mt$ & $\mathcal{O}(N_\theta^2 )$ & $\mathcal{O}(N_\theta^2 )$ \\ 
& $\force$ & $\mathcal{O}(N_\theta N_H)$ & $\mathcal{O}(N_\theta C_H )$\\ 
&  & \\ 
\hline
\multirow{3}{4em}{OVQITE} &  &  \\ 
& $M$ & $\mathcal{O} (N_\theta |S_\tau|)$ & $\mathcal{O}(N_\theta C_{S_\tau})$\\ 
& $\bm v$ & $\mathcal{O}(N_H |S_\tau|)$ & $\mathcal{O}(C_{\{H,\hat O\},H,S_\tau})$ \\ 
&  &  \\ 
\hline
\end{tabular}
\caption{Number of quantum circuits required for VQITE and OVQITE. For OVQITE, $G$ and $\bm b$ are functions of $M$ and $\bm v$, see Eq.~\eqref{eq-G-from-M} and~\eqref{eq-b-from-M-v}. $N^{(\mathrm{naive})}_{\mathrm{circ}}$ is the number of circuit to be prepared when considering all terms to be measured individually, whilst $N^{(\mathrm{opt})}_{\mathrm{circ}}$ corresponds to the number of simultaneous measurements of qubit-wise commuting groups.}
\label{table}
\end{table}

\section{RESULTS}\label{sec:results}
In this section, we present the numerical results of the classical simulation of a  quantum hardware implementation of VQITE and OVQITE, and we analyze numerically the convergence to the the ground state of the tranverse-field Ising model (TFIM) as a function of the total number of measurements per step. 

\subsection{The transverse-field Ising model}
We implement the OVQITE algorithm for the one-dimensional TFIM with periodic boundary conditions, whose Hamiltonian is given by:
\begin{equation}
		\label{tfim}
		\hat{H}=-J\sum_i \hat{Z}_i \hat{Z}_{i+1}-h\sum_i\hat{X}_i,
	\end{equation} 
where $J>0$ is the Ising interaction strength between nearest-neighbor spins and $h$ is the strength of the external magnetic field along the $x$ direction. In this work, we benchmark the ground state preparation in two regimes: in the ordered ferromagnetic phase at $h/J=1/2$ and in the critical phase at $h/J=1$ where the system undergoes a quantum phase transition to the disordered phase. 

\subsection{Quantum circuit ansatz}
\begin{figure}[t!]
      \centering
    \includegraphics[width=\linewidth]{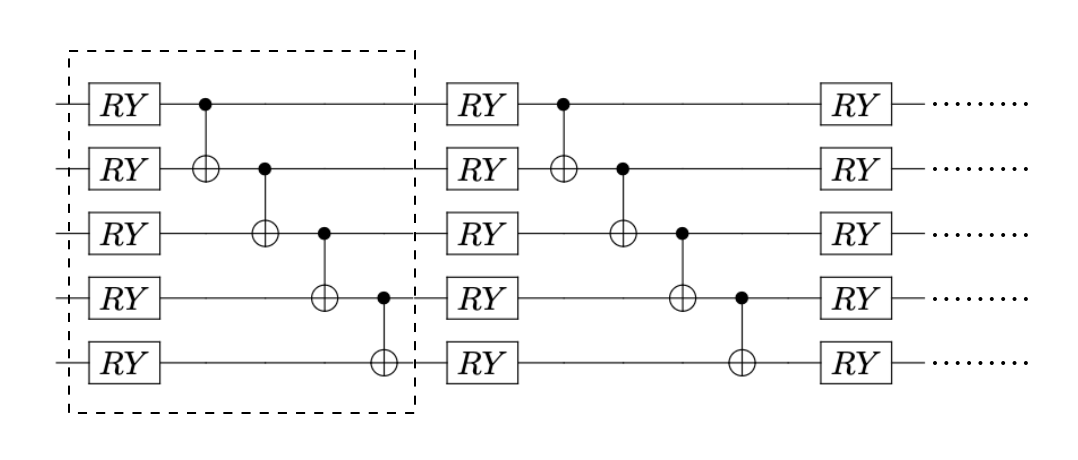}
      \caption{Hardware efficient ansatz (HEA) quantum circuit for $n=5$ qubits used in the simulation of the VQITE and OVQITE algorithms in this work. One layer is composed of $R_y(\theta)$ rotations and a set of CNOT gates between neighboring qubits arranged in a staircase pattern. The dashed square indicates one layer of the ansatz.}
      \label{fig:ansatz}
\end{figure}

For the preparation of the variational state, we use a hardware efficient ansatz (HEA)~\cite{vqe1}, which consists of a sequence of $L$ layers of parameterized Pauli rotation gates $R_Y(\theta) \equiv e^{i\theta Y}$, followed by CNOT gates arranged in staircase pattern, see Fig.~\ref{fig:ansatz}.  For $n$ qubits, the total number of tunable parameters is $N_\theta=n(L+1)$.

\subsection{Choice of the operator set $S_{\tau}$}
Let us now discuss the choice of the operator set $S_\tau$ used in our numerical implementation of OVQITE. It is easy to verify that if $\hat{H}\in\mathrm{Span}\,S_\tau$, then the family of density matrices $\hat\eta_\tau$ which solve Eq.~\eqref{eq-ite-projected-operators} either converges to an eigenstate of $\hat H$, or the average energy decreases as a function of $\tau$. This indicates that $\hat H \in \mathrm{Span}\,S_\tau$ is a natural requirement for the operator set $S_\tau$. We will designate this minimal choice as the 
Hamiltonian set as $S_\tau = \setH$, where
\begin{equation}\label{eq:operator_set_H}
\setH \equiv\{\hat{H}\}.
\end{equation}
For the TFIM Hamiltonian of Eq.~\eqref{tfim}, $|\setH|=2\,n$.

Another natural choice for $S_\tau$ is, as discussed in Sec.~\ref{subsec:oite}, the set of all 1- and 2-local operators. For efficiency reasons, we consider in this work a subset $S_\tau=\setNN$ of all 1- and 2-local operators containing only Pauli strings including nearest-neighbor Pauli operators,
\begin{equation} \label{eq:operator_set_NN}
\setNN\equiv\bigcup_{\alpha, j}\{\hat{P}_{\alpha;j}\}\bigcup_{\alpha,\gamma,\braket{j,k}}\{\hat{P}_{\alpha;j}\hat{P}_{\gamma;k}\},
\end{equation}
where $\hat{P}_{\alpha;j}$ is the $\alpha$-th Pauli matrix acting on the qubit at lattice site $j$, and $\braket{j,k}$ means that the sites $j$ and $k$ are nearest neighbors on the lattice. For Hamiltonians that correspond to a sum of 1-local and 2-local nearest-neighbor terms, one has $\hat H \in \mathrm{Span}\,\setNN$. For the TFIM Hamiltonian of Eq.~\eqref{tfim}, $|\setNN|=12\,n$.

As a further simplification of the $\setNN$ set of operators in the case of a real quantum circuit such as the HEA we use in this work, see Fig.~\ref{fig:ansatz}, we can consider $S_\tau=\setTFIM$ 
\begin{equation}
    \setTFIM \equiv \setNN \setminus\bigcup_{j}  \{\hat Y_j\}\setminus\bigcup_{j,k}  \{\hat Y_j\hat X_k,\hat Y_j\hat Z_k\},
\end{equation}
which coincides with the set $\setNN$ after removing the imaginary Pauli strings that have zero expectation value. This leads to a reduced factor in the number of operators, $|\setTFIM|=7\,n$.

We additionally note that there is no restriction on adding $k$-local operators with $k>2$ to the operator set and this indeed would be desired for other systems of interest (i.e. from quantum chemistry). In the limit of adding all possible operators to the set one recovers the VQITE result.

 \begin{figure}
      \centering      \includegraphics[width=\linewidth]{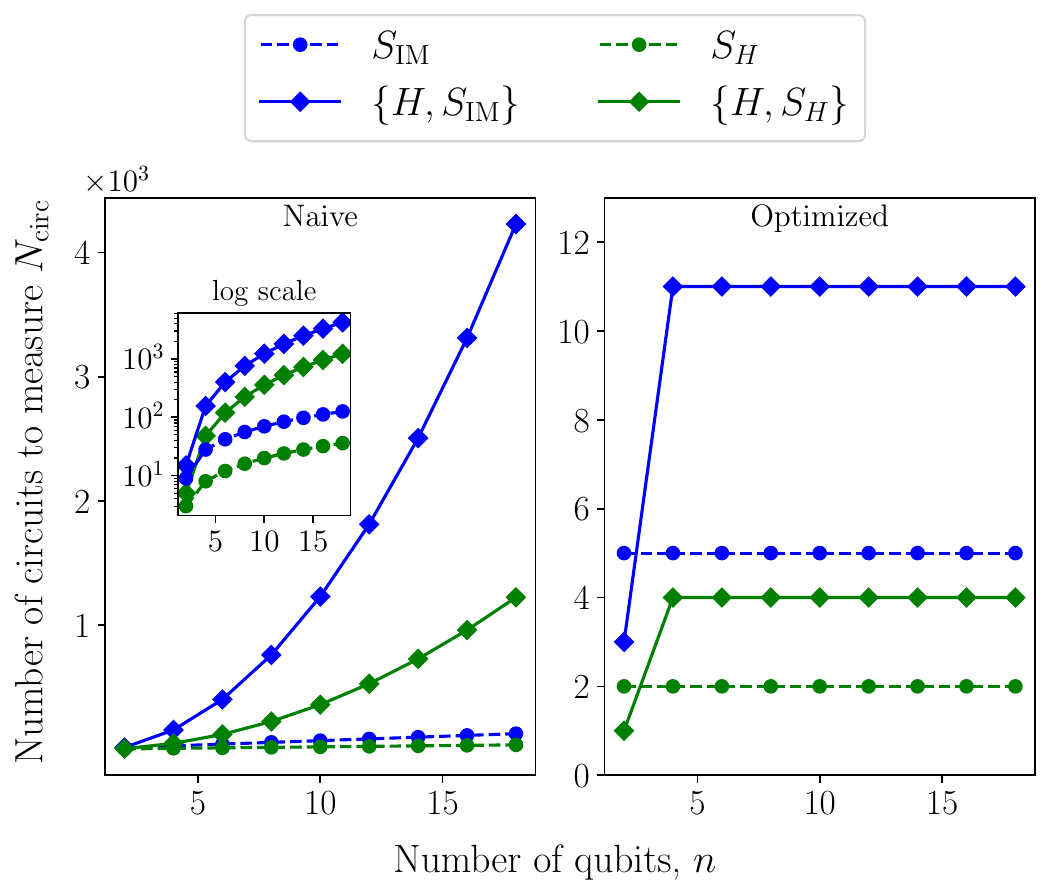}
      \caption{Number circuits to be measured within one step of OVQITE as a function system size $n$ of the TFIM. We show results for the operator sets $S_{\text{IM}}$ and $S_{H}$ (dashed blue and green lines, respectively) as well as their anticommutators with the Hamiltonian (solid lines). \emph{Left:} All operators are measured individually. The inset shows the same data on a logarithmic scale. \emph{Right:}  Qubit-wise commuting operators are grouped and measured simultaneously.}
      \label{fig:term}
\end{figure}

 \begin{figure}
      \centering
\includegraphics[width=\linewidth]{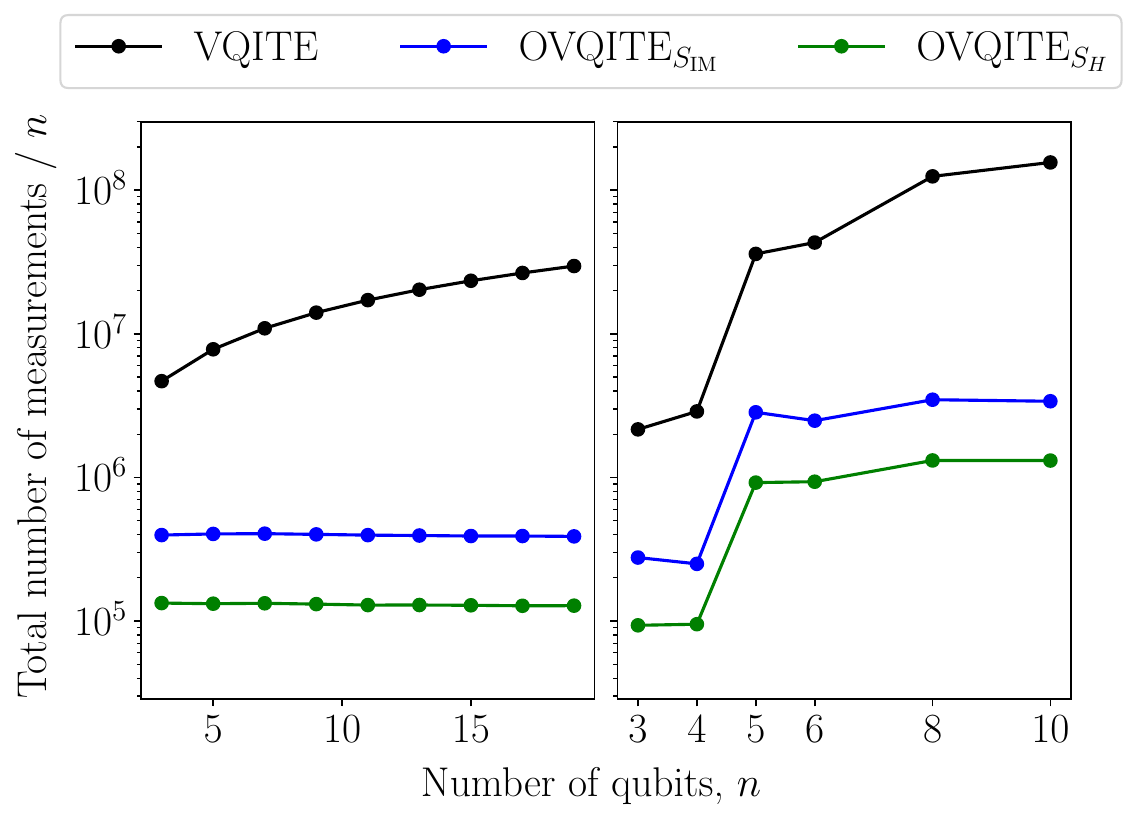}
\caption{Scaling analysis of VQITE and OVQITE for the TFIM. \emph{Left:} Theoretical scaling of the total number of measurements per iteration normalized by the system size $n$. The number of HEA layers in the parameterized circuit is fixed to $L=5$ and $N_\mathrm{sh}=10^4$ shots are used per expectation value or overlap calculation. \emph{Right:} Experimental numbers of measurements  at which a relative energy error accuracy of $\Delta E/E=10^{-3}$ is reached using an optimized measurement strategy as described in Appendix~\ref{sec:meas}. }\label{fig:meas}
\end{figure}

\subsection{Number of measurements}
In Fig.~\ref{fig:term} we show the dependence of the number of circuits on the number of qubits required for the implementation of the OVQITE algorithm for the TFIM. Specifically, we show the number of circuits needed to measure all the elements of the two operator sets $S_\tau=\setH$ and $S_\tau=\setTFIM$, as well as their respective anticommutators with the Hamiltonian operator. From the left panel of Fig.~\ref{fig:term} we observe that, without grouping, the number of terms in the anticommutators is the bottleneck of both choices of $S_\tau$, scaling quadratically as a function of the number of qubits $n$. As shown in the right panel of Fig.~\ref{fig:term}, if one considers the simultaneous measurement of the expectation values for groups of operators which qubit-wise commute~\cite{McClean_2016, meas1, vqe}, the total number of groups is small and independent of the system size for both operator sets. We provide further details of our grouping strategy in Appendix~\ref{sec:meas}.

\begin{figure}
      \centering
\includegraphics[width=\linewidth]{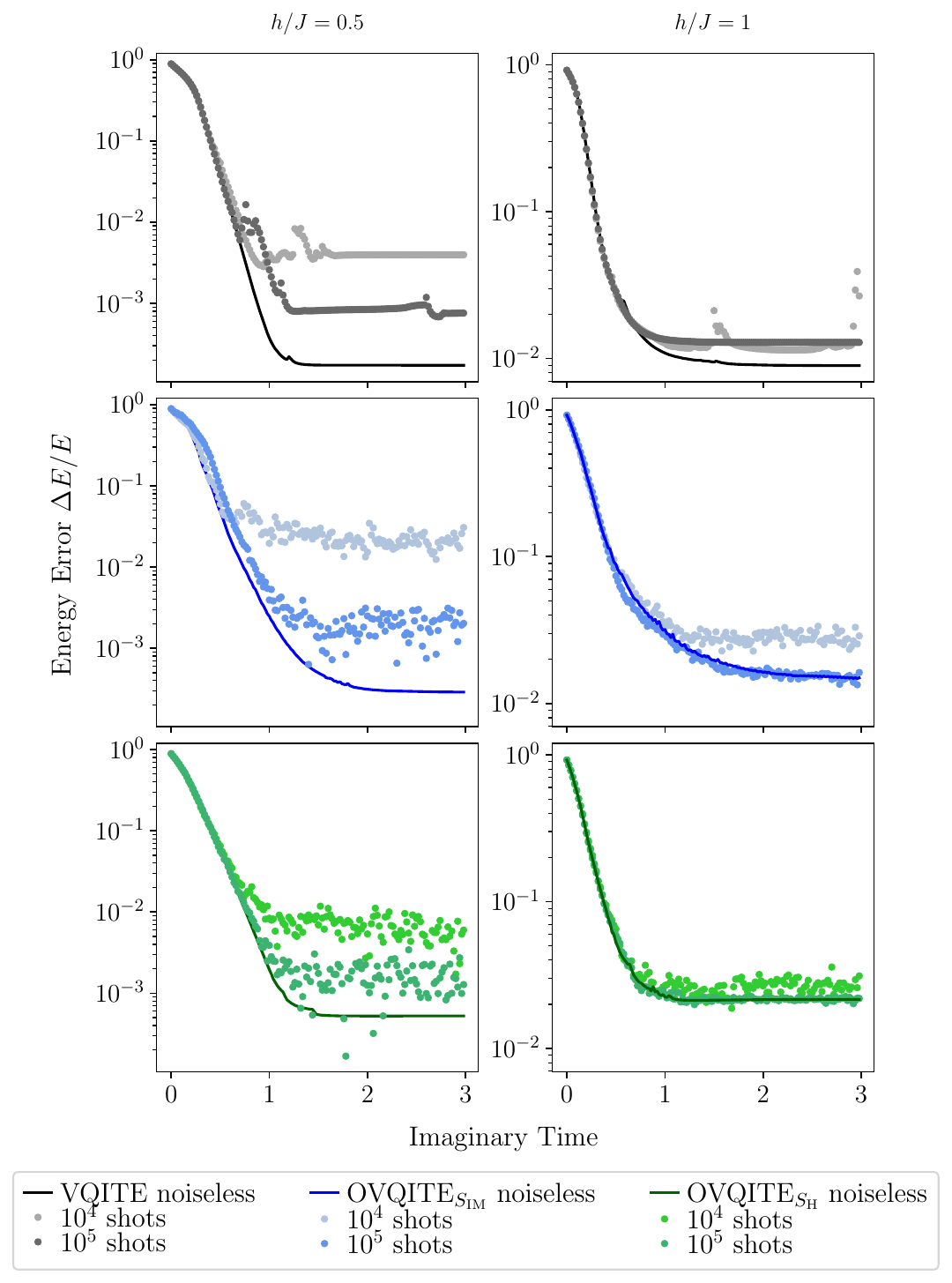}

      \caption{The convergence of the relative error on the  ground state energy as a function of imaginary time. Results are shown for OVQITE$_{\setH}$ (green), OVQITE$_{\setTFIM}$ (blue) and VQITE (black) for the 10-qubit TFIM with a $L=5$ layer HEA ansatz. Noiseless imaginary-time evolution is represented by solid curves, whilst results from $N_{\text{sh}}=10^4$ and $N_{\text{sh}}=10^5$ shots (see Eq.~\eqref{eqn-shots}) are shown as data points. Two TFIM regimes are studied, $h/J=0.5$ (left) and $h/J=1$ (right).}
      \label{fig:all}
\end{figure}

In Fig.~\ref{fig:meas} we present the scaling of VQITE, OVQITE$_{S_{H}}$ and OVQITE$_{S_{\text{IM}}}$ in terms of the total number of measurements per iteration, normalized by the system size $n$. The left panel of Fig.~\ref{fig:meas} shows the theoretical scaling of total number of measurements computed using Table \ref{table} as we fix the number of HEA circuit layers to $L=5$ and the number of shots per expectation value estimation to $N_{\mathrm{sh}}=10^4$. We see that the ratio of measurements to system size saturates for both OVQITE implementations, while it grows linearly for VQITE. In the right panel of Fig.~\ref{fig:meas} we show the numerical confirmation of this scaling for system sizes of up to $n=10$ qubits. Here, we use a measurement strategy as defined in Appendix~\ref{sec:meas} to reach a target accuracy on the relative energy error of $\Delta E/E= 10^{-3}$.  We remark that the overall scaling of the three algorithms  follows the theoretical prediction coming for the number of circuits $N_{\mathrm{circ}}$ of Sec.~\ref{sec-scaling-analysis}. We further note that the  number  of measurements for VQITE is already two orders of magnitude higher than OVQITE at the investigated system sizes. Between the two OVQITE implementations, OVQITE$_{\setH}$ is more efficient than OVQITE$_{\setTFIM}$ at a given accuracy, which is expected since the corresponding operator set is smaller.

\begin{figure}
      \centering
\includegraphics[width=\linewidth]{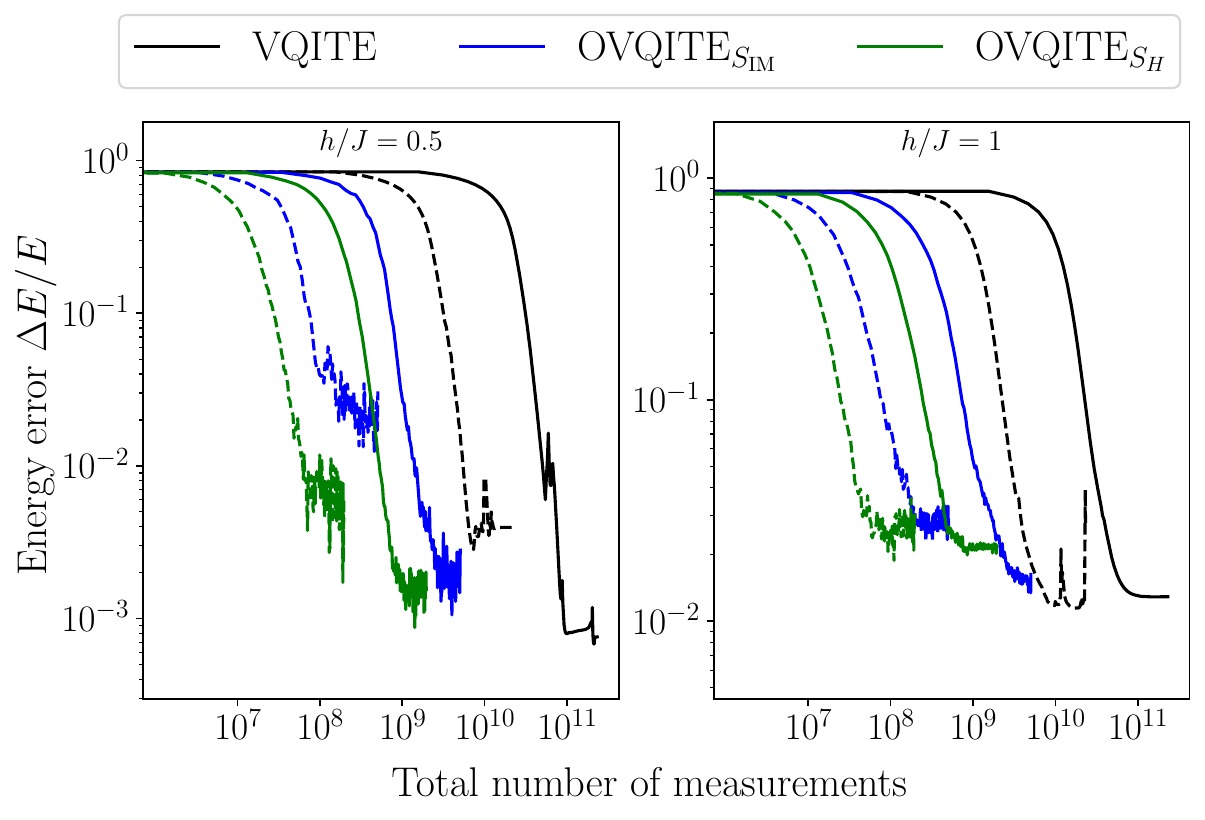}
\caption{Energy relative error on the ground state energy for the 10-qubit TFIM at  $h/J=0.5$ (left) and $h/J=1$ (right) is shown as a function of the number of measurements for OVQITE$_{S_H}$ (green), OVQITE$_{S_{\text{IM}}}$ (blue) and VQITE (black). The influence of shot noise is investigated for $N_{\text{sh}}=10^4$ (dashed lines) and $N_{\text{sh}}=10^5$ (solid lines).} \label{fig:meas_all}
\end{figure}

\subsection{Convergence in the presence of shot noise}

 Next, we test the convergence for the three algorithms with and without sampling noise. For the latter, we use Qiskit Sampler primitives~\cite{qiskit1} to simulate the circuit with $N_{\text{sh}}=10^4$ and $N_{\text{sh}}=10^5$ shots used to estimate each expectation value for OVQITE or overlap for VQITE. We initiate all three algorithms from random parameters and we use Eq.~\eqref{eq-s-matrix-and-f} with $\delta =0.02$ for $h/J=0.5$ and $\delta =0.015$ for $h/J=1$. 
 Details of the regularization procedure are given in Appendix~\ref{sec:pinv}.

In Fig.~\ref{fig:all} we show the convergence of the average energy to the ground state value for the three algorithms within $150$ imaginary time steps. In the absence of shot noise, the three methods are able to reach the ground state within an accuracy of up to $\Delta E/ E \sim 10^{-3}$ at $h/J=0.5$ (left panels) and $\Delta E/ E \sim 10^{-2}$ for the more challenging regime at $h/J=1$ (right panels). A clear hierarchy is observed, where the computationally cheaper methods converge to higher energy errors.

When shot noise is taken into account, we observe the expected deterioration of the accuracy of all algorithms. For $h/J=0.5$, the accuracy for $10^5$ shots drops to around $\Delta E / E \sim 5\times 10^{-3}$  for OVQITE$_{S_\text{IM}}$ and OVQITE$_{S_{H}}$ and to around $\Delta E/ E \sim 10^{-3}$ for VQITE. With only $10^4$ shots it further decreases to around $\Delta E/ E \sim 10^{-2}$ for VQITE and OVQITE$_{S_{H}}$.
%
%
From investigating the $h/J=1$ regime (right panels) it is clear that the results are much less affected by shot noise. This can be explained by the already significantly less accurate energies in the absence of it, which means that the number of available samples per expectation value limits the accuracy to the same values irrespective of the overall accuracy of the algorithms in a given regime. It can be thus concluded that the required number of samples can be roughly inferred solely from the desired accuracy of the result.

Let us now turn our focus to Fig.~\ref{fig:meas_all}, where we examine the number of measurements required to obtain a given accuracy. For all three algorithms, the error decreases exponentially and monotonically until a certain accuracy threshold is reached, after which it abruptly slows down and starts to oscillate, reminiscent of Fig.~\ref{fig:all}. In the exponentially decreasing regime, the computationally cheaper method always fares better, with OVQITE$_{S_H}$ requiring about $5\times$ less measurements than OVQITE$_{S_\text{IM}}$ and around $100\times$ less compared to VQITE for the same number of samples per expectation value or overlap, $N_{\text{sh}}$. This allows, e.g., to reach an accuracy of $\Delta E / E \sim 5\times 10^{-2}$  for $h/J=0.5$ in under $10^8$ total measurements.

We can also compare the performance of the methods between $h/J=0.5$ and $h/J=1$ by looking at accuracy $\Delta E / E \sim 5\times 10^{-1}$, which is achieved by all algorithms and for both $N_{\text{sh}}=10^4$ and $N_{\text{sh}}=10^5$. Interestingly, for any given algorithm and number of samples, roughly the same accuracy is reached in both regimes. We conclude that, generally, the cheapest OVQITE method should be preferentially used as long as it stays within the exponentially improving regime. To this end, it is easy to identify the moment this is no longer the case, at least when not considering the effects of hardware noise. Once the limit of this regime is reached, it is, to some degree, possible to mitigate this effect by increasing the number of shots per expectation value. Once this no longer works, one may consider switching between different operator sets or changing the complexity of the ansatz circuit. We leave such adaptive approaches to future studies.

\section{Discussion}\label{sec:conclusions}

We have proposed a new variational quantum algorithm to simulate imaginary time evolution projected onto a set of operators and benchmarked it on the ground state preparation of the transverse-field Ising model Hamiltonian. Compared to the established VQITE method~\cite{mclahans}, the algorithm brings a number of advantages. 

First, by avoiding the need of estimating quantum fidelities, the required circuits are only half in length, which could lead to exponentially improved overall performance when using noisy quantum hardware. 

Second, the scaling of the total number of measurements with the number of variational parameters is reduced from quadratic to linear, as it eliminates the need for the evaluation of the Quantum Fisher Information matrix. As a consequence, OVQITE requires about two orders of magnitude less measurements to achieve a similar accuracy to VQITE.  However, the measurement complexity remains model dependent since the number of terms in the operator set of OVQITE will be a function of the number of terms in the Hamiltonian of interest, and the scaling will depend on how terms from the anticommutator of Eq.~\ref{eqn:V} are split into qubit-wise commuting groups.

We have further studied the influence of measurements noise with  $10^4$ and $10^5$ shots per evaluation on the convergence of the three algorithms and found that this convergence is independent on the parameter regime, but rather a function of the target accuracy. We also found that it depends only weakly on the choice of quantum imaginary-time evolution algorithm, as for all three an accuracy of at least $5\times10^{-2}$ and $1\times10^{-2}$ was achieved for $10^4$ and $10^5$ shots in all parameter regimes, respectively. We found that, especially at critical point of the TFIM ($h/J=1$), there was a fundamental limit to the achievable accuracy of any of the three algorithms, something that could potentially be improved by evolving for longer times and using deeper and more expressive ansatz circuits. To this end, it would be worthwhile to study the impact of the structure of the ansatz circuit on the overall performance of OVQITE. 

The results of this work have been obtained with an optimized regularization procedure based on the pseudo-inverse method and all three algorithms could be stabilized by introducing a cut-off for small singular values in the linear problem. Consequently, this procedure introduces a trade-off between the stability and accuracy of the aforementioned algorithms. Various other regularization techniques could be explored in the future, in particular when taking into account shot noise. In this paper, we have proposed a new approach (detailed in Appendix~\ref{sec:error-in-variables}) that updates parameters in the direction that maximizes the likelihood of solving the linear problem when stochastic shot noise errors are present.

Whilst we have managed to significantly improve the scaling of the number of measurements for performing imaginary time evolution on a quantum processor, even the $\sim10^{8}$ shots required for a relative error of $10^{-2}$ are close to the limit of what is executable within a reasonable time-frame (hours to days) on currently available quantum devices. This demands further improvement in the technique for it to become applicable to a broader and more complex class of problems. We believe additional performance boosts can be readily achieved in a number of ways. First, techniques should be developed for the identification of optimal operator sets, whether with respect to the  accuracy or the measurement overhead. These operator sets could, in principle, be generated adaptively or stochastically, as we have seen that smaller sets are sufficient for reaching lower accuracies. Second, the number of shots per observable can equally be changed adaptively to achieve optimal statistical estimates as defined by the maximum likelihood principle of Appendix~\ref{sec:error-in-variables}. Third, we have only considered constant imaginary-time steps in this work, yet these can also be altered in the course of the algorithm execution, and be determined from the rate of change of relevant observables.  Finally, the total number of measurements could be further reduced by the use of more advanced measurement and grouping methods, such as, e.g., the shadow grouping scheme developed in Ref.~\cite{gresch2023guaranteed}.

So far, we have only studied the performance of OVQITE for quantum spin systems. It would be of interest to also apply our method to fermionic systems from condensed matter and quantum chemistry. The same is true for the impact of real quantum hardware noise on the performance of OVQITE, which we intend to investigate in future studies.

\section{Acknowledgements}
The authors would like to thank G. Carleo, A. Kahn, M. Leib and F. Vicentini for useful discussions. RR ackowledges financial support from ``Appel à projets au fil de l’eau du Centre d’Information Quantique'', Sorbonne Universit\'e, and from SEFRI through Grant No. MB22.00051 (NEQS - Neural Quantum Simulation).
\let\l\originalpolishl

\bibliography{main}
\clearpage
\onecolumngrid

\let\l\redefinedl

\section{APPENDIX}

\label{sec:pinv}
\subsection{Regularization}
The goal is to solve a system of linear equations $Ax=b$. If $A$ is invertible, it has an exact solution $x=A^{-1}v$. Near singular values, the solution becomes unstable as the reciprocal of a small number results in amplification of errors. We instead solve the system using the pseudo-inverse with an SVD decomposition. The observation matrix can be decomposed as $A=U\Sigma V^T$ where $\Sigma$ is a diagonal matrix containing singular values $\sigma_i$ and $U,V$ are orthogonal matrices.
To regularize the instability, we impose a condition on the reciprocal of singular values $\sigma'$ such as 
\begin{equation}
\label{eqn:redefining_sing_values}
    \sigma'_i= 
\begin{cases}
    \frac{1}{\sigma_i},& \text{if } \sigma_i\geq \text{threshold}\\
    0,              & \text{if } \sigma_i < \text{threshold}
\end{cases}
\end{equation}
where we define 
\begin{equation}
    \text{threshold}=\sigma_{\text{max}} \times \textbf{rcond}
\end{equation}
such that $\sigma_\text{max}$ is the largest singular value and $\textbf{rcond}$ is a small auxiliary parameter. 
Eq.~\eqref{eqn:redefining_sing_values} sets those singular values $\sigma_i$ that are smaller than the threshold to zero and takes the reciprocal of the remaining ones, which allows the inversion process to be stabilized. 
This regularized pseudo-inverse retains the essential information from the original matrix $A$ while discarding the components that could lead to numerical instability.
Then one can reconstruct the observation matrix as
\begin{equation}
    A_\text{pinv}=V\Sigma^{'-1}U^T
\end{equation}
with $\Sigma^{'-1}=\text{diag}(\sigma'_1,\sigma'_2,...,\sigma'_n)$. The regularized solution can hence be computed from $x=A_{pinv}b$. \\

Table \ref{table:reg} summarizes the optimal regularization parameter that stabilizes the three different algorithms; VQITE, OVQITE$_{S_\mathrm{IM}}$,OVQITE$_{S_H}$ for a for $10$ qubits TFIM. We show results for the exact computation of eigenvalues and eigenvectors of the Hamiltonian (left), and simulations involving shot noise (right).  

\begin{table}[h!]
\begin{center}
\begin{tabular}{ |c|c|c|c| } 
\hline
h/J & Optimizer & Rcond (exact) & Rcond (with shot noise) \\
\hline
\multirow{3}{*}{$0.5$}& VQITE & $10^{-6}$ & $10^{-3}$\\
    &OVQITE$_{S_{\mathrm{IM}}}$ & $10^{-5}$ & $10^{-4}$\\
    &OVQITE$_{S_{H}}$ & $10^{-4}$ & $10^{-4}$\\
    \hline\hline

\multirow{3}{*}{$1$}& VQITE & $10^{-6}$ & $10^{-3}$\\
    &OVQITE$_{S_{\mathrm{IM}}}$ & $5\times 10^{-6}$ & $5\times 10^{-5}$\\
    &OVQITE$_{S_{H}}$ & $10^{-4}$ & $10^{-4}$\\
    \hline
\end{tabular}
\end{center}
\caption{Regularization parameters used for VQITE and OVQITE for imaginary-time evolving the $n=10$ qubit TFIM in different parameter regimes.}
\label{table:reg}
\end{table}
\FloatBarrier

 \subsection{Optimized measurements}
 \label{sec:meas}
 If two operator commute $[A,B]=0$, we can determine a common eigenbasis $|\phi\rangle$ that simultaneously diagonalizes both $A$ and $B$. Their expectation value can be computed as : 

\begin{equation}
\label{eqn:exp}
\langle A\rangle =\langle \psi|\big(\sum_n\lambda_{A,n}|\phi_n\rangle\langle \phi_n|\big)|\psi\rangle=\sum_n\lambda_{A,n}|\langle \phi_n|\psi\rangle|^2,
\end{equation}
and
\begin{equation}
\label{eqn:exp2}
\langle B\rangle =\langle \psi|\big(\sum_n\lambda_{B,n}|\phi_n\rangle\langle \phi_n|\big)|\psi\rangle=\sum_n\lambda_{B,n}|\langle \phi_n|\psi\rangle|^2,
\end{equation}

The following algorithm computes $|\langle\phi_n|\psi\rangle|^2$ for each group of Pauli operators that qubit-wise commute. The eigenvalues $\lambda_i$ can be obtained by computing the Kronecker product of eigenvalues of Pauli matrices in a given Pauli string. Here, we will use qubit-wise commutativity where two Pauli strings $P_i$ and $P_j$ qubit-wise commute if for each qubit $k$ we have $[P_i^k,P_j^k]=0, \; \forall k$.

\begin{algorithm}[H]
\caption{Compute Expectation Value of Commuting Pauli Strings}
\begin{algorithmic}[1]
    \Require List of Pauli strings $\{P_1, P_2, \ldots, P_n\}$

    \State Initialize an empty list of groups $\{G_1, G_2, \ldots, G_m\}$
    
    \For{each Pauli string $P$ in the list}
        \State Initialize a flag $placed \gets \text{False}$
        \For{each group $G_k$}
            \If{$P$ commutes qubit-wise with every Pauli string in $G_k$}
                \State Add $P$ to $G_k$
                \State $placed \gets \text{True}$
                \State \textbf{break}
            \EndIf
        \EndFor
        \If{not $placed$}
            \State Create a new group $G_{m+1} \gets \{P\}$
            \State Append $G_{m+1}$ to the list of groups
        \EndIf
    \EndFor

    \For{each group $G_k$}
        \State Pick a representative Pauli string $P_k$ from $G_k$
        \For{each qubit $i$ in $P_k$}
            \If{$P_k^i = X$}
                \State Apply Hadamard gate $H$ on qubit $i$
            \ElsIf{$P_k^i = Y$}
                \State Apply S-gate followed by Hadamard gate $SH$ on qubit $i$
            \EndIf
        \EndFor
        \State Measure all qubits in the computational basis
        \State Record measurement results
    \EndFor
    
    \State Classically post-process expectation values using recorded results using Equation \ref{eqn:exp},\ref{eqn:exp2}
\end{algorithmic}
\end{algorithm}

\FloatBarrier


\subsection{Solving linear problems with error in variables}
\label{sec:error-in-variables}
\paragraph{\textbf{The setup.}}
We consider the problem 
\begin{equation}
\label{eqn:problem}
B = A^* x + \varepsilon \, ,
\quad 
A = A^* + \delta \, 
\end{equation}
where $(B,A)$ are the observed variables, $B$ being a vector and $A$ a matrix, distributed as Gaussians $\cN(B^*,\Omega_B)$ and $\cN(A^*,\Omega_A)$ respectively, 
where $B^* = A^* \beta$.
(Equivalently, the errors are supposed to be distributed as $\varepsilon \sim \cN(0,\Omega_B)$ and $\delta\sim \cN(0,\Omega_A)$.)
The covariance matrix $\Omega_B$ and covariance tensor $\Omega_A$ are supposed known, while the means $B^*$ and $A^*$ are not.

We search for the best estimator of $x$, which is linear in the known parameters.
Here, ``best'' is in terms of the maximum likelihood principle, i.e.\ we want to maximize the probability $P(B,A\vert x)$ of having the observed $B$ and $A$ in our (single) measurement.
This estimator for $x$ will be referred to as \emph{maximum likelihood estimator}, or MLE in short.



\medskip

\paragraph{\textbf{The distribution of the difference.}}
Let us consider the variable
$$
D = B-Ax \, .
$$
Being sum of Gaussian distributions, $D$ is known to be a Gaussian distribution (given $x$). 
It has mean conditional to $x$ given by
\[
\EE_x[D]= \EE_x[B-Ax]=\EE_x[D]-\EE_x[A]x = B^* - A^*x = 0
\]
and covariance matrix
\begin{align*}
\EE_x[(D-\EE_x[D]) (D-\EE_x[D])^T] =&
\EE_x\left\{[B-B^*-(A-A^*)x][B-B^*-(A-A^*)x]^T\right\} \\
 =&
\EE_x[(B-B^*)(B-B^*)^T] - \EE_x[(B-B^*)\beta^T(A-A^*)^T] \\
&
- \EE_x[(A-A^*)\beta (B-B^*)^T] + \EE_x[(A-A^*)xx^T(A-A^*)^T] \, .
\end{align*}

Now, let's assume that the conditional covariances $\Covar_x(A_{i,j},B_k)$ between the components $A_{i,j}$ of $A$ and $B_k$ of $B$ are zero, for all $i,j,k$.
Then, 
\[
(\EE_x[(B-B^*)x^T(A-A^*)^T])_{i,j} = \Covar_x(B_i,\sum_m A_{j,m}x_m) = 
\sum_m x_m \, \Covar_x(B_i, A_{j,m}) = 0 \, ;
\]
and similarly $\EE_x[(A-A^*)\beta (B-B^*)^T] = 0$.

We then also define $\Omega_B$ a matrix with elements $(\Omega_B)_{i}^j$ given by the conditional covariance $\Covar_x(B_i,B_j)$ of the components $B_i$ with $B_j$ of $B$. 
Similarly, we denote by $\Omega_X\equiv \Var_x(X)$ the tensor with components $(\Omega_A)_{i,m}^{j,l}\equiv \Covar_x(A_{i,m},A_{j,l})$.
Then,
\[
(\EE_x[(A-A^*)xx^T(A-A^*)^T])_{i,j} = 
\Covar_x((\sum_m A_{i,m}x_m)(\sum_l A_{j,l}x_l)) = 
\sum_{m,l} x_m x_l \Covar_x(A_{i,m},A_{j,l}) = (x^T \Omega_A x)_{i,j} \, ,
\]
where in the last term the contraction with $x^T$ on the left is intended on the index $m$, whereas the contraction with $\beta$ on the right on the index $l$. (Note that, if helpful for numerical implementations, the tensor contraction $x^T\Omega_A x$ can be transformed to a product of matrices.)

Then, we simply get 
\[
\Omega_D\equiv \Var_x(D) = \Omega_B + x^T \Omega_A x \, .
\]
In conclusion, we have conditional law
\[
D \sim_x \cN(0,\Omega_B + x^T\Omega_A x) \, ,
\]
i.e.\
\[
P(D=d \, \vert x ) = \frac{e^{\, d^T(\Omega_B + x^T\Omega_A x)^{-1}d}}{\sqrt{(2\pi)^n\det(\Omega_B + x^T\Omega_A x )}}
\]
where $n$ is the size of $D$.

\medskip

\paragraph{\textbf{Gradient descent for the the maximum likelihood estimation.}}

Eliminating the terms which do not depend on $x$, which are irrelevant as far as the maximum likelihood estimator for $x$ is concerned, we can compute
\begin{align*}
P(A=a,B=b \, |\, x)&=
P(A=a|x) P(B=b|A=a,x)\\
&\propto  P(B=b|A=a,x) \\
&=P(D=b-ax\, |\, A=a,x)\\
&=P(D = b - ax \, |\, x) \, ,
\end{align*}
in other words 
\begin{equation}
\label{eqn:prob}
P(A=a,B=b \vert x ) \propto \frac{e^{\, (b-ax)^T(\Omega_Y + x^T\Omega_A x)^{-1}(b-ax)}}{\sqrt{\det(\Omega_B + x^T\Omega_A x)}} \, .
\end{equation}

The maximum likelihood estimator $\hat{x}$ (or MLE in short) of $x$ is then the minimizer of the quantity $P(A=a,B=b\vert x)$ above.
As the latter has a rather complicated dependence on $x$, we propose to apply (inverse) gradient descent starting from $x=0$. 
This approach is justified by the empiric idea that the point of maximum closest to $x=0$ is the actual global minimum of the function, which is true in the limit of $A$ not stochastic.
To this end, we explicitly compute the gradient of the right hand side of \eqref{eqn:prob}.

\bigskip

We start by recalling the following derivation rules:
\begin{enumerate}
    \item Given a $\mathbb{R}$-parametric family of invertible matrices $M_t$, we have
    \[
    \frac{\d}{\d t} M_t^{-1} = - M^{-1}_t \cdot \frac{\d M_t}{\d t} \cdot M_t^{-1} \, .
    \]
    
    \item Given a $\mathbb{R}$-parametric family of invertible matrices $M_t$, we have
    \[
    \frac{\d}{\d t}\det(M_t) =
    \det(M_t) \, \tr(M_t^{-1}\frac{\d M_t}{\d t})
    \]
\end{enumerate}
Using these rules, one can compute
\begin{equation}
\label{eqn:gradient}
\begin{split}
    \frac{\d}{\d x_s} P(A=a,B=b\vert x) \;\propto & \;
    \frac{    e^{\, (b-ax)^T(\Omega_Y + x^T\Omega_A x)^{-1}(b-a x)}}{
    \det(\Omega_B + x^T\Omega_A x)
    }
    \, \\
    &
    \left\{
    \det(\Omega_B + x^T\Omega_A x) \, \tr\left((\Omega_B + x^T\Omega_A x)^{-1}\frac{\d (x^T\Omega_A x)}{\d x_s}\right)
    \right. 
    \\
    &
    +
    \sqrt{\det(\Omega_B + x^T\Omega_A x)}
    \left[
    \frac{\d (ax)^T}{\d x_s} (\Omega_B + x^T\Omega_A x)^{-1}(b-a x) \right.
    \\
    &
    - (b-a x)^T 
     (\Omega_B + x^T\Omega_A x)^{-1}  \frac{\d (x^T\Omega_A x)}{\d x_s}  (\Omega_B + x^T\Omega_A x)^{-1}
     (b-a x)
    \\
    &
    \left.\left.
    + 
    (b-a x)^T(\Omega_B + x^T\Omega_A x)^{-1} \frac{\d (a x)}{\d x_s}
    \right]\right\}
\end{split}
\end{equation}

More explicitly, the derivatives in the formula above read as follows in components:
\[
    \left(\frac{\d (a x) }{\d x_s}\right)_{i} = a_{i,s} 
\]    
\[
\left(\frac{\d (x^T\Omega_A x)}{\d x_s} \right)_{i,j} = \sum_{l}x_l \Covar_x(A_{i,s},A_{j,l}) + 
    \sum_{m}x_m \Covar_x(A_{i,m},A_{j,s})
\]

\paragraph{\textbf{Regularization}}

If a regularization is needed, it can be done by imposing a prior distribution.
More precisely, what said in the previous sections remains true, as only probability conditional to $\beta$ were used.

Then, instead of maximizing $P(X=x,Y=y\vert\beta)$, we now aim to maximize 
\[
P(A=a,B=b) = P(A=a,B=b \,\vert \, x)\, P(x) \, ,
\]
which using a prior $x \sim \cN(0,\lambda \Id)$, for some arbitrary small regularization parameter $\lambda>0$, just becomes
\begin{equation}
\begin{split}
\label{eqn:prob_regularized}
P(A=a,B=b )  &\propto \frac{e^{\, (b-a x)^T(\Omega_B + x^T\Omega_A x)^{-1}(b-a x)}}{\sqrt{\det(\Omega_B + x^T\Omega_A x)}} \,\cdot\, \frac{e^{\, \frac{\Vert x\Vert^2}{\lambda}}}{\sqrt{\lambda}} \\
& = \frac{e^{\, (b-a x)^T(\Omega_B + x^T\Omega_A x)^{-1}(b-a x) + \frac{x^Tx}{\lambda}}}{\sqrt{\det(\Omega_B + x^T\Omega_A x)}}
\, .
\end{split}
\end{equation}

\bigskip

The gradient descent approach above can then be performed with this new function in \eqref{eqn:prob_regularized}, which amounts to use the following instead of \eqref{eqn:gradient}:

\begin{equation}
\label{eqn:gradient_regularized}
\begin{split}
    \frac{\d}{\d x_s} P(A=a,B=b\vert x) \;\propto & \;
    \frac{    e^{\, (b-a x)^T(\Omega_B + x^T\Omega_A x)^{-1}(b-a x)+ \frac{x^Tx}{\lambda}}}{
    \det(\Omega_B + x^T\Omega_A x)
    }
    \, \\
    &
    \left\{
    \det(\Omega_B + x^T\Omega_A x) \, \tr\left((\Omega_B + x^T\Omega_A x)^{-1}\frac{\d (x^T\Omega_A x)}{\d x_s}\right)
    \right. 
    \\
    &
    +
    \sqrt{\det(\Omega_B + x^T\Omega_A x)}
    \left[
    \frac{\d (ax)^T}{\d x_s} (\Omega_B + x^T\Omega_A x)^{-1}(b-a x) \right.
    \\
    &
    - (b-a x)^T 
     (\Omega_B + x^T\Omega_A x)^{-1}  \frac{\d (x^T\Omega_A x)}{\d x_s}  (\Omega_B + x^T\Omega_A x)^{-1}
     (b-ax)
    \\
    &
    \left.\left.
    + 
    (b-a x)^T(\Omega_B + x^T\Omega_A x)^{-1} \frac{\d (ax)}{\d x_s} \;
    +\;  \frac{2x_s}{\lambda}
    \right]\right\} \, .
\end{split}
\end{equation}

\end{document}